\documentclass[12pt]{article}
\usepackage{graphicx}
\usepackage{float}
\usepackage{amssymb}
\usepackage{amsthm}
\usepackage{amstext}
\usepackage{amsmath}
\usepackage{feynmp-auto}
\usepackage{caption}
\usepackage{subcaption}
\usepackage{xfrac}
\usepackage[hidelinks]{hyperref}

\newcommand\pubdate{\today}

\def\institute{Nikhef, Science Park 105, Amsterdam, The Netherlands}

\def\Title#1{\begin{center} {\Large #1 } \end{center}}
\def\Author#1{\begin{center}{ \sc #1} \end{center}}
\def\Address#1{\begin{center}{ \it #1} \end{center}}

\newcommand\pubblock{\rightline{\begin{tabular}{l} \pubdate  \end{tabular}}}
\newenvironment{Abstract}{\begin{quotation}  }{\end{quotation}}
\newenvironment{Presented}{\begin{quotation} \begin{center} 
             PRESENTED AT\end{center}\bigskip 
      \begin{center}\begin{large}}{\end{large}\end{center} \end{quotation}}

\def\beq{\begin{equation}}
\def\eeq#1{\label{#1}\end{equation}}
\def\eeqn{\end{equation}}

\def\beqa{\begin{eqnarray}}
\def\eeqa#1{\label{#1}\end{eqnarray}}
\def\eeqan{\end{eqnarray}}

\let\bar=\overbar

\def\Dslash{\not{\hbox{\kern-4pt $D$}}}
\def\dslash{\not{\hbox{\kern-2pt $\del$}}}

\def\msb{{\bar{\ssstyle M \kern -1pt S}}}

\newcommand{\FDFI}{\left(\varphi^\dagger\overleftrightarrow{D}^I_\mu\varphi\right)}

\begin{document}
\begin{titlepage}
\pubblock

\vfill
\Title{Dimension six effective operators in $t$-channel single top production and decay at NLO in QCD}
\vfill
\Author{ Marc de Beurs }
\Address{\institute}
\vfill
\begin{Abstract}
I summarize phenomenological LHC13 studies of the effect of dimension six effective operators on the inclusive cross section and differential distributions for the t-channel single top process at NLO in QCD~\cite{thePaper}.
\end{Abstract}
\vfill
\begin{Presented}
$11^\mathrm{th}$ International Workshop on Top Quark Physics\\
Bad Neuenahr, Germany, September 16--21, 2018
\end{Presented}
\vfill
\end{titlepage}
\def\thefootnote{\fnsymbol{footnote}}
\setcounter{footnote}{0}

\section{Standard Model effective theory and single top quark production}

With the large amounts of data that the LHC is delivering we are well inside the era of precision physics. This makes it possible to not only stress-test the Standard Model, but also to measure or constrain new physics. The framework of Standard Model Effective Field Theory allows us to describe these unknown effects in a model independent way, while being able to incorporate the symmetries of the SM.\\

\noindent Since top quarks are singly produced via the weak charged current interaction, the relevant production and decay vertex are the same, as can be seen in Figure~\ref{fig:diagram}. This narrows down the number of effective operators that can enter in the process. It also means that the width of the top quark can be affected. Moreover, an interference between effective operators in the production and the decay vertex can occur. Finally, single top quarks produced in this way are polarized which makes the study of additional angular observables interesting.\\



\noindent We extend the SM Lagrangian by including the relevant dimension-6 operators ${\cal O}_i$, with their associated coefficient $C_i$. The scale of new physics $\Lambda$ is set to 1 TeV.

\begin{equation}
  {\cal L}_{\mathrm{SM}} + \sum_i  \frac{C_i }{\Lambda^2} O^{[6]}_i +{\mathrm{hermitian\; conjugate}}
\end{equation}

\noindent Only the following three operators can enter our process at tree level at ${\cal O} \left ( \sfrac{1}{\Lambda^2} \right )$

\begin{minipage}[T]{0.55\textwidth}
\begin{eqnarray}
  O_{\varphi Q}^{(3)} &=&i \frac{1}{2} y_t^2 \FDFI (\bar{Q}\gamma^\mu\tau^I Q) \label{O1}\\
  O_{tW}&=&y_t g_w(\bar{Q}\sigma^{\mu\nu}\tau^It)\tilde{\varphi}W_{\mu\nu}^I \label{O2}  \\
  O^{(3)}_{qQ,rs}&=&\left(\bar q_r\gamma^\mu \tau^I q_s\right)\left(\bar Q\gamma_\mu \tau^I Q\right)  \label{O3}
\end{eqnarray}
\end{minipage}~
\begin{minipage}[T]{0.4\textwidth}
\begin{figure}[H]
\centering
\begin{fmffile}{diagram}
\resizebox{5cm}{!}{
\begin{fmfgraph*}(200,200)
\fmfleft{i1,i2,i3,i4,i5,i6}
\fmfright{o1,o2,o3,o4,o5,o6}
\fmf{fermion,label=$b$,tension=1}{i2,v1}
\fmf{fermion,label=$t$}{v1,v3}
\fmf{zigzag,label=$W$,tension=1.5}{v1,v2}
\fmfv{d.sh=circle,d.fi=full, d.si=3thick, f=(1,,0,,0)}{v1}
\fmf{fermion,tension=1.5}{i5,v2}
\fmf{fermion}{v2,o5}
\fmf{zigzag,label=$W$}{v3,o2}
\fmf{fermion,label=$b$}{v3,o3}
\fmfv{d.sh=circle,d.fi=full, d.si=3thick, f=(1,,0,,0)}{v3}
\end{fmfgraph*}
}
\end{fmffile}
\write18{mpost diagram}
\caption{Diagram for the $t$-channel single top production and decay process.}
\label{fig:diagram}
\end{figure}
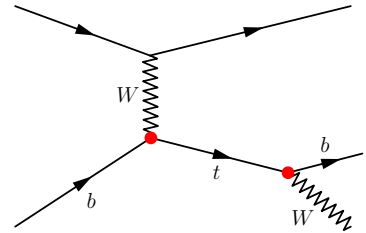
\end{minipage}

\noindent where we use the same notation as in Ref.~\cite{Zhang:2016omx}.

\section{Cross-section results}

We consider the process $p p \to W b j$ (decayed top) for which the computation is performed using {\sc MadGraph5\_aMC@NLO} \cite{Alwall:2014hca}. The NLO EFT implementation of
Ref.~\cite{Zhang:2016omx} makes use of the 5 flavour scheme (b quarks are massless). The $W$ boson is decayed leptonically through {\sc MadSpin} \cite{Artoisenet:2012st},
and {\sc Pythia8} \cite{Sjostrand:2014zea} is used for parton
showering and hadronisation. Since we also generate the irreducible
backgrounds, a loose invariant mass cut is imposed on the $Wb$ system,
centred on the top mass $100\, \mathrm{GeV} < M_{Wb-\mathrm{jet}} < 250\, \mathrm{GeV}$ 
\cite{Frederix:2016rdc}.\\

\noindent The choice for our benchmark coefficients of the effective operators can be found in Table~\ref{tab:cvalues}, together with the corresponding cross section and width of the top quark. The predicted deviations from the SM lie within the uncertainty of recent single top measurements: $\sigma = 156 \pm 35$ pb and
$0.6 \le \Gamma_{\text{top}} \le 2.5$ GeV~\cite{CMS:2016hdd,Aaboud:2016ymp}\\

\begin{table}[H]
\renewcommand{\arraystretch}{1.8}
 \makebox[\linewidth]{
  \small
	\begin{tabular}{l|c|cc|cc}
& & \multicolumn{2}{|c|}{LO} & \multicolumn{2}{|c|}{NLO}\\
Operator & Coupling value & $\sigma$[pb] $\pm$scale $\pm$PDF & $\Gamma_{\text{top}}$ [GeV] & $\sigma$[pb] $\pm$scale $\pm$PDF & $\Gamma_{\text{top}}$ [GeV] \\
\hline
$\text{SM}$ &  - & $123^{+9.3\text{\%}}_{-11.4\text{\%}} \pm 8.9\text{\%}$ & 1.49 & $137^{+2.7\text{\%}}_{-2.6\text{\%}} \pm{1.2\text{\%}}$ & 1.36\\
$O^{(3)}_{qQ,rs}$ & -0.4 & $172^{+8.7\text{\%}}_{-10.8\text{\%}} \pm 8.9\text{\%}$ & 1.49 & $190^{+2.4\text{\%}}_{-1.8\text{\%}} \pm 1.1\text{\%}$ & 1.35 \\
$O_{\varphi Q}^{(3)}$ & 1 & $137^{+9.3\text{\%}}_{-11.4\text{\%}} \pm 8.9\text{\%}$ & 1.67 & $154^{+2.3\text{\%}}_{-2.3\text{\%}} \pm 1.2\text{\%}$ & 1.52\\
$O_{tW}$ (Re) & 2 & $132^{+9.3\text{\%}}_{-11.4\text{\%}} \pm 8.8\text{\%}$ & 1.83 & $148^{+2.3\text{\%}}_{-2.5\text{\%}} \pm 1.2\text{\%}$ & 1.68\\
$O_{tW}$ (Im) & 1.75i & $125^{+9.2\text{\%}}_{-11.4\text{\%}} \pm 8.8\text{\%}$ & 1.51 & $140^{+2.3\text{\%}}_{-2.5\text{\%}} \pm 1.2\text{\%}$ & 1.38\\
\end{tabular}}
\caption{The benchmark choices for the coefficients of the effective operators, together with the corresponding $t$-channel single top cross section and the width of the top quark. The scale and PDF uncertainties of the cross sections are also shown.}
\label{tab:cvalues}
\end{table}

\section{Polarization angles}

The polarisation angle $\theta_i^z$ is the angle between the
direction of decay product i and the spectator jet, as viewed in the top rest frame.
The angular distribution of any top decay product in this frame can be parametrised as
\begin{equation}
  \frac{1}{\sigma}\frac{d\sigma}{d \cos\theta^z_i}=\frac{1}{2}\left(1+a_i P\, \cos\theta^z_i \right)
  \label{eq:pol_angle} 
\end{equation}
where $P$ denotes the top quark polarisation and $a_i$ encodes how
much spin information is transferred to each decay product. For the
charged lepton $a_l$ is close to 1, indicating nearly 100\%
correlation.\\

\noindent We use the same reference system as in \cite{Aguilar-Saavedra:2014eqa} to construct a new set of coordinates:
\begin{equation}
  \hat{z} = \frac{\vec{p_j}}{|\vec{p_j}|}, \;\;\;\;\;\; \hat{y} = \frac{\vec{p_j} \times \vec{p_q}}{|\vec{p_j} \times \vec{p_q}|}, \;\;\;\;\;\; \hat{x} = \hat{y} \times \hat{z}\,.
\label{eq:pol-3D}
\end{equation}
The vectors $\vec{p}_j$ and $\vec{p}_q$ represent the direction of the spectator- and of the
initial quark, respectively,  both in the top quark rest-frame.
Since the initial quark cannot be known with certainty, the beam axis is used.\\

\noindent We investigate the distributions of the angles between the directions of
the top quark decay products and these new directions. The angle of
the charged lepton with respect to the three axes defined above is affected mostly by the polarisation of the top \cite{Mahlon:2000ze}.\\

\noindent It should be noted that we have exclusively studied Monte Carlo distributions. No background processes are taken into account, apart from the SM contributions to Wbj production and only minimal selection criteria that resemble the ATLAS detector acceptance are applied: leptons must have a $p_T$ of at least 10 GeV and lie in the region $|\eta| < 2.47$, whereas jets must have a $p_T$ bigger then 20 GeV and that lie inside $|\eta| < 4.5$.

\section{Sensitivity}

\begin{figure}[H]
\centering
\begin{subfigure}{0.3\textwidth}
\centering
\includegraphics[scale=0.24,page=7]{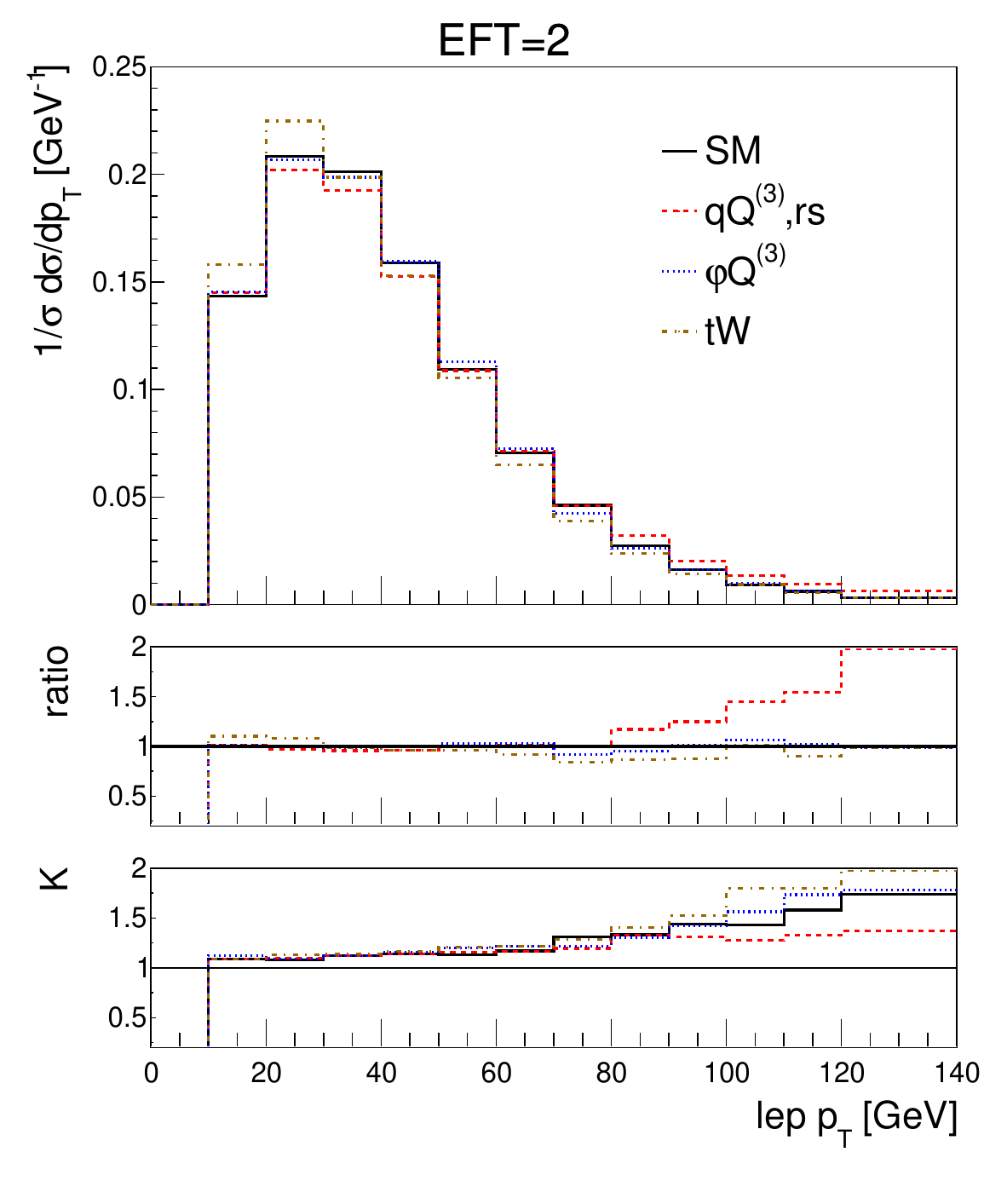}  
\end{subfigure}%
\begin{subfigure}{0.3\textwidth}
\centering
\includegraphics[scale=0.24,page=11]{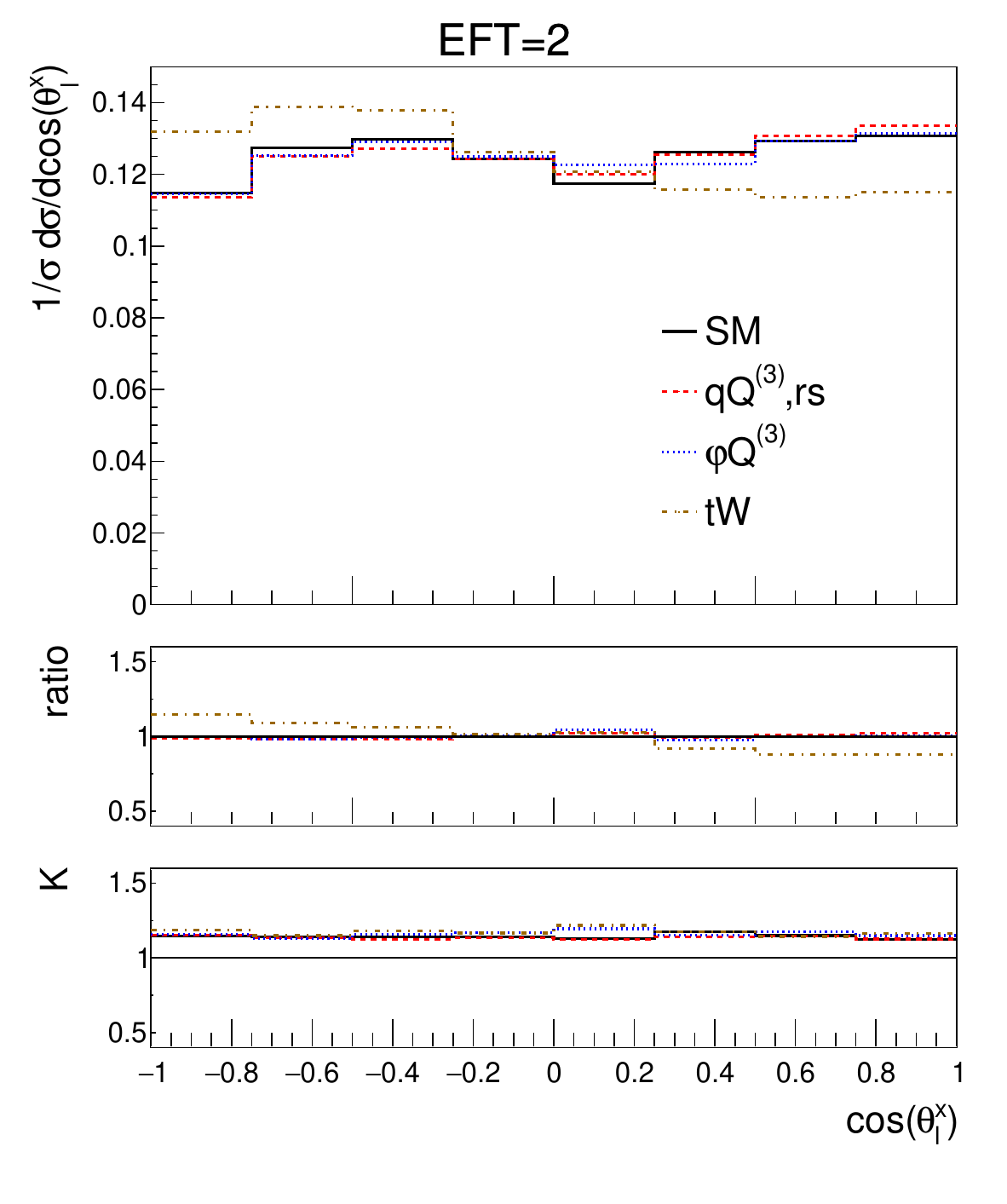}
\end{subfigure}%
\begin{subfigure}{0.3\textwidth}
\centering
\includegraphics[scale=0.24,page=7]{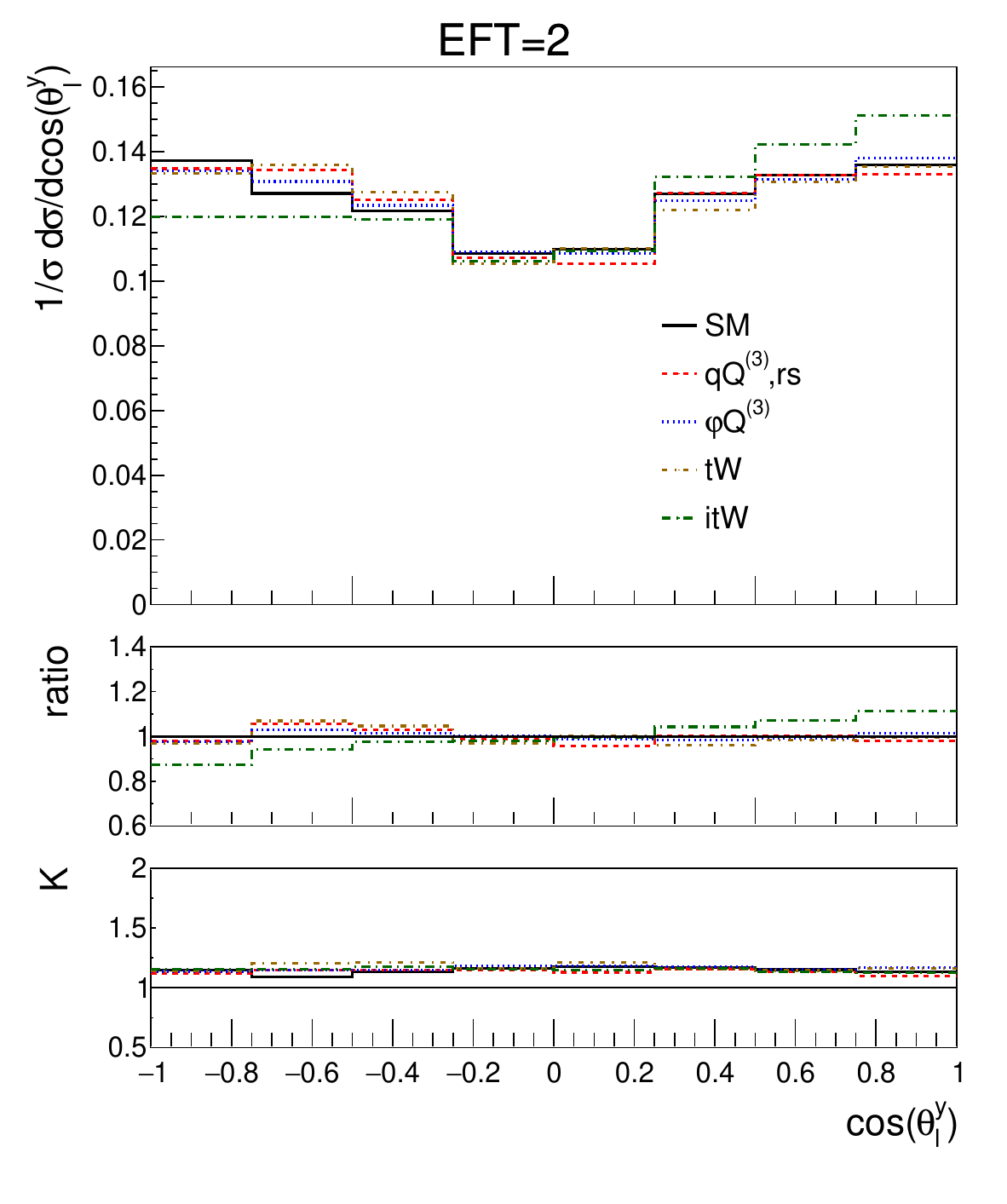}  
\end{subfigure}
\caption{The NLO distributions of the lepton transverse momentum (left) and two of the three polarization angles, the $\theta_i^x$ (middle) and $\theta_i^y$ (right). The coefficients of the effective operators can be found in Table \ref{tab:cvalues}. The ratio shown in the first inset is defined as the effect of the operator over the SM, the second inset shows the
  $K$-factor which is defined as the ratio of the NLO over the LO predictions.}
\label{fig:sensitivity}
\end{figure}

Figure~\ref{fig:sensitivity} shows that different distributions are sensitive to different operators, which makes it possible to distinguish between operators. The 4-fermion operator ($O_{qQ^{(3)},rs}$) leads to harder leptons (left). The lepton $p_T$ is a good observable to measure in experiments. Angles are also an accessible observable in measurements, especially since one could increase the sensitivity by measuring the asymmetry between forward and backward scattering. In the plot on the right we show that one can also probe the imaginary part of the dipole operator ($O_{tW}$), which reveals the presence of CP violation.\\
 
\noindent It is important to notice that the ratios of the NLO over LO predictions ($K$-factors) are rather large, and that the distributions show that there is a difference in shape for the lepton $p_T$ between LO and NLO, it is not just a normalization.  

\section{Conclusion}

In this work we computed for the first time single top
production and decay at NLO in QCD, in the presence of dimension-6
operators. QCD corrections are found to affect both the total rates and the differential distributions in a non-trivial way.\\

\noindent We find that different distributions are sensitive to different operators and that an angular distribution can be used to identify CP-violating effects coming from the imaginary 
part of the dipole operator coefficient.\\

\noindent This work has been done in collaboration with Eleni Vryonidou, Eric Laenen and Marcel Vreeswijk. The full study can be found at~\cite{thePaper}.

\end{document}